\documentstyle[prl,aps,amsfonts,amssymb,epsfig,floats]{revtex}
\begin{document}
\draft
\preprint{}
%
%following two lines are for two column format
\twocolumn[\hsize\textwidth\columnwidth\hsize\csname@twocolumnfalse\endcsname
\title{Superconducting Gap and Strong In-Plane Anisotropy in Untwinned YBa$_2$Cu$_3$O$_{7-\delta}$}
\author{D. H. Lu, D. L. Feng, N. P. Armitage, K. M. Shen, A. Damascelli, C. Kim, F. Ronning, and Z.-X. Shen}
\address{Department of Physics, Applied Physics and Stanford Synchrotron Radiation Laboratory, \\Stanford University, Stanford, CA 94305, USA}
\author{D. A. Bonn, R. Liang, and W. N. Hardy}
\address{Department of Physics and Astronomy, University of British Columbia, Vancouver, BC V6T 1Z1, Canada}
\author{A. I. Rykov and S. Tajima}
\address{Superconductivity Research Laboratory, International Superconductivity Technology Center, \\1-10-13, Shinonome, Koto-ku, Tokyo 135-0062, Japan}

\date{Received 23 January 2001}
\maketitle

\begin{abstract}
With significantly improved sample quality and instrumental resolution, we
clearly identify in the ($\pi$,0) ARPES spectra from YBa$_2$Cu$_3$O$_{6.993}$,
in the superconducting state, the long-sought `peak-dip-hump' structure. This
advance allows us to investigate the large $a$-$b$ anisotropy of the in-plane
electronic structure including, in particular, a 50\% difference in the
magnitude of the superconducting gap that scales with the energy position of
the hump feature. This anisotropy, likely induced by the presence of the CuO
chains, raises serious questions about attempts to quantitatively explain the
YBa$_2$Cu$_3$O$_{7-\delta}$ data from various experiments using models based on
a perfectly square lattice.
\end{abstract}
%
%74.25.Jb Electronic structure
%74.72.Bk Y-based cuprates
%79.60.Bm Clean metal, semiconductor, and insulator surfaces
\vspace{-0.3cm}

\vskip2pc]
\narrowtext

High-temperature superconductivity (HTSC) is intimately related to the CuO$_2$
plane, which is the only common structural feature in all cuprates. This fact
has led most of the proposed microscopic theories to assume a CuO$_2$ square
planar structure. However, for the practical reason of sample quality, some of
the most important and defining experiments have been performed on
YBa$_2$Cu$_3$O$_{7-\delta}$ (Y123), which does not have a square lattice, but
rather an orthorhombic structure ($b/a\approx 1.015$), caused by the presence
of a CuO chain layer \cite{jorgenson87}. This orthorhombicity, according to LDA
calculation \cite{anderson95}, should result in significant anisotropy in the
{\it in-plane} electronic structure (this term will be used throughout this
paper to refer to the electronic states associated with the CuO$_2$ plane),
making it problematic to compare theories based on a square lattice with
experimental data from Y123. Therefore, it is crucial to quantify the effect of
orthorhombicity, if any, on the in-plane electronic structure in Y123. The
problem is that angle-resolved photoemission spectroscopy (ARPES), being a
uniquely powerful tool for this important task, has, until now, not been
particularly effective for the study of Y123 \cite{shen95}. The important
`peak-dip-hump' structure, which is seen routinely in
Bi$_2$Sr$_2$CaCu$_2$O$_{8+\delta}$ (Bi2212) \cite{dessau91} has never been
observed in Y123. This absence, together with the presence of a surface state
\cite{schabel98}, raises questions about ARPES data from Y123, and the
universality of the superconducting peak in the cuprates.

This paper reports a breakthrough in this important issue, made possible by
significantly improved sample quality and instrumental resolution. By isolating
a surface state peak near the Fermi energy ($E_F$), we can clearly resolve a
`peak-dip-hump' structure in the ARPES spectra around ($\pi,0$) in Y123 that
resembles the superconducting peak observed in Bi2212 \cite{dessau91}. More
significantly, we find a strong $a$-$b$ asymmetry of the in-plane electronic
structure, such as the superconducting gap magnitude, which differs by about
50\%. We argue that such a strong in-plane $a$-$b$ anisotropy should be taken
into account when interpreting experiments performed on Y123.

ARPES experiments were carried out at beamline 5-4 at SSRL, which is equipped
with a normal-incidence-monochromator and a SCIENTA SES-200 analyzer. Untwinned
YBa$_2$Cu$_3$O$_{6.993}$ single crystals (overdoped, $T_c$ = 89 K) with
superior chemical purity ($99.99-99.995\%$) and crystallinity (FWHM =
0.007$^{\circ}$ in the X-ray rocking curve) were grown by the self-flux method
in BaZrO$_3$ crucibles \cite{liang98}. Single crystals were cleaved {\it
in-situ} at 10 K with a base pressure better than 5x10$^{-11}$ {\it torr}.
ARPES spectra were recorded at a photon energy of 28 eV with an energy
resolution of 10 meV and an angular resolution of 0.3$^{\circ}$, corresponding
to a momentum resolution of 1.5\% of the Brillouin zone (BZ). Low energy
electron diffraction (LEED) patterns were routinely acquired at the end of
experiments to check the quality of the cleaved surfaces and to confirm the
sample orientation.

\begin{figure}[t!]
\centerline{\epsfig{figure=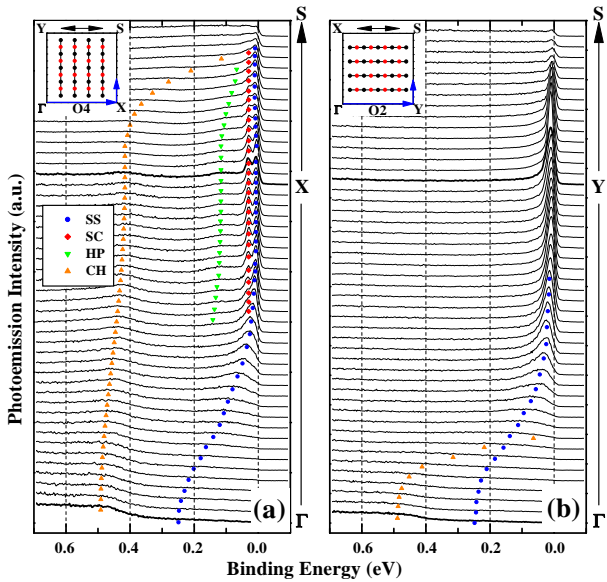,width=8.0cm,clip=}} \vspace{0.2cm}
\caption{(color) ARPES spectra taken at 10 K on untwinned
YBa$_2$Cu$_3$O$_{6.993}$ along (a) $\Gamma$-X-S and (b) $\Gamma$-Y-S. The
insets depict the sample orientations (black and red circles represent the CuO
chains) with respect to the linear photon polarization (black arrows). Four
features in the spectra, labeled as SS, SC, HP, and CH, correspond to surface
state, superconducting peak, hump, and chain state, respectively (see text).}
\label{edc}
\end{figure}

\begin{figure}[t!]
\centerline{\epsfig{figure=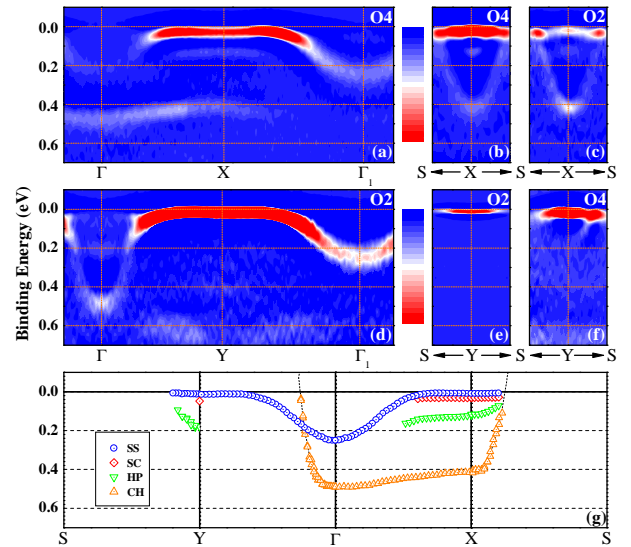,width=8.0cm,clip=}} \vspace{0.2cm}
\caption{(color) (a)-(f) Second derivatives of the ARPES spectra along high
symmetry lines with different sample orientations. The red-white-blue color
scale is chosen to emphasize the broad features that are difficult to resolve
in the EDCs. Bottom panel (g) summarizes all of the band dispersions.}
\label{disp}
\end{figure}

Fig.\ \ref{edc} presents the energy distribution curves (EDCs) along high
symmetry directions in the BZ. Data were taken at 10 K under two distinct
sample orientations with respect to the polarization of the incoming radiation
(insets), which are referred to as O4 and O2, following the existing notation
\cite{schabel98}. In order to show the dispersion clearly, in particular for
the broad and weak features, the intensity plots of the second derivatives of
the EDCs are displayed in Fig.\ \ref{disp}a-\ref{disp}f. Four features can be
clearly identified, two of which appear at X as sharp peaks close to $E_F$,
along with two broad features at higher binding energy (BE). As we will justify
in the following paragraphs, we assign them, from the lower to higher BE, as
surface state, superconducting peak, hump, and chain state. Fig.\ \ref{disp}g
summarizes all of the band dispersions.

The most pronounced feature in Fig.\ \ref{edc} is a narrow, intense peak just
below $E_F$ at both X and Y, whose high sensitivity to surface degradation at
elevated temperatures indicates its surface state character. The precise origin
of this surface state is, however, not well understood at present. One
possibility could be the surface termination effect. An earlier scanning
tunneling microscopy (STM) study \cite{edwards92} demonstrated that Y123
cleaves between the CuO chain and BaO layers, resulting in both CuO
chain-terminated and BaO-terminated areas on the cleaved surface. The scenario
of a chain-related surface state is favored by a previous ARPES study
\cite{schabel98}, and seems to be consistent with more recent STM data
\cite{derro00}. However, this assignment is difficult to reconcile with the
two-dimensional dispersion of this feature along $\Gamma$-X and $\Gamma$-Y, as
seen in the current data. For this paper, as we will focus on the bulk
electronic structure, we leave it as an open question.

For the broad feature located well below $E_F$ at X, instead of a bonding
$\sigma$ state as suggested in Ref.\ \cite{schabel98}, we reassign it to a
chain derived state. This assignment is concluded from the quasi
one-dimensional (1D) character of this feature, {\it i.e.}, the dispersion is
strong along the chain direction ($\Gamma$-Y and X-S) and weak in the
perpendicular direction ($\Gamma$-X). Further support comes from the fact that
the two crossings along $\Gamma$-Y and X-S agree very well with the 1D chain
Fermi surface (FS) determined by a full FS mapping. The dispersion of this
chain state and the related chain FS are consistent with the LDA band structure
\cite{anderson95}. It is also interesting to note that this feature looks
rather similar to the chain state observed in PrBa$_2$Cu$_3$O$_7$
\cite{mizokawa99} and that the dispersion along the chain direction
qualitatively agrees with the holon band predicted by the $t$-$J$ model
calculation at quarter filling \cite{maekawa96}. However, we were not able to
identify in the current data a spinon band, which may possibly be masked by the
hump around 0.12 eV in Y123.

\begin{figure}[t!]
\centerline{\epsfig{figure=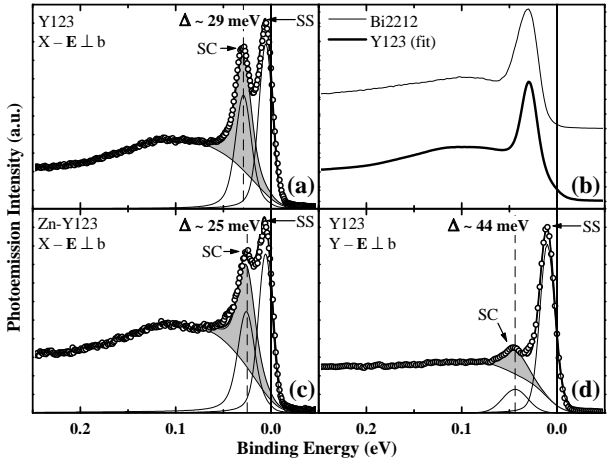,width=8.0cm,clip=}} \vspace{0.2cm}
\caption{(a) EDC at X reproduced from Fig.\ \ref{edc}a together with fitting
curves. Panel (b) compares the fitting curve (after subtracting the surface
state peak) with the EDC from overdoped Bi2212 ($T_c$ = 84 K) at ($\pi,0$). (c)
EDC at X from Zn-doped Y123 with fitting curves. (d) EDC at Y from Y123 with
fitting curves. All data shown were taken at 10 K. Shaded area represents
weight of the superconducting peak.} \label{scp}
\end{figure}

Now let us focus on the superconducting peak and the hump that are clearly
evident in Fig.\ \ref{edc}a. For better illustration, the EDC at X is replotted
in Fig.\ \ref{scp}a with fitting curves. In the fitting procedure, an
integration background that smoothly merges into the high-energy hump is
subtracted to extract the double peak feature, which is then fitted with a
simple spectral function: a product of a Fermi function and two Lorentzians,
convolved with a gaussian instrumental resolution function. After subtracting
the surface state peak, the rest of the spectrum looks strikingly similar to
the well-known `peak-dip-hump' structure observed in the ARPES spectra of
Bi2212 at ($\pi,0$) below $T_c$ (Fig.\ \ref{scp}b), indicating that the second
peak may be a superconducting peak.

\begin{figure}[t!]
\centerline{\epsfig{figure=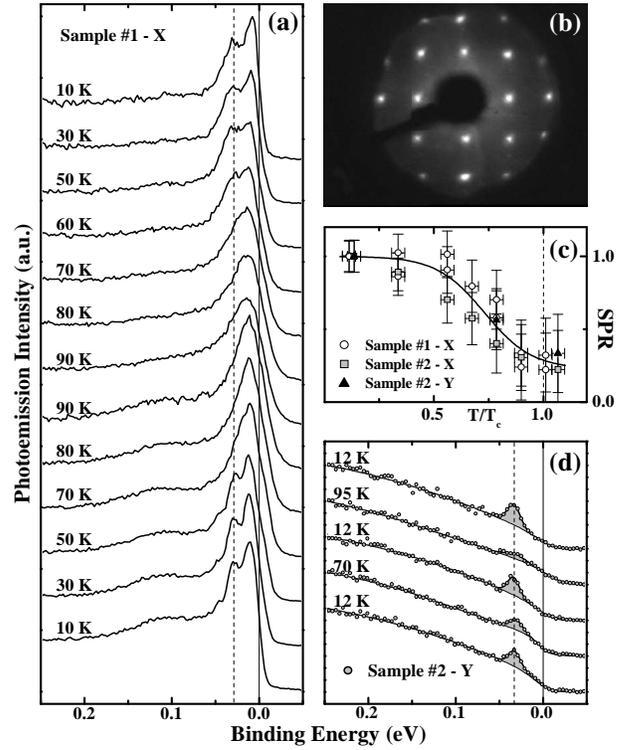,width=8.0cm,clip=}} \vspace{0.2cm}
\caption{Temperature dependence of ARPES spectra from Y123 at (a) X-point and
(d) Y-point. Data are presented in the order of acquisition starting from the
bottom curve of each panel. (b) LEED pattern taken after experiments for panel
(a). (c) Normalized SPR (see text) for different samples {\it vs}. reduced
temperature.} \label{temp}
\end{figure}

More direct evidence of the superconducting nature of this peak comes from its
temperature dependence (Fig.\ \ref{temp}a). Due to the unstable nature of the
Y123 surface at high temperatures, great caution has been taken in cycling the
temperature. A LEED pattern taken after the measurements (Fig.\ \ref{temp}b)
shows no visible difference from that of a freshly cleaved surface, indicating
a well-ordered surface, although some minor aging effects can be detected.
Following the same fitting procedure (with both peak positions fixed), we
extract the superconducting peak and plot in Fig.\ \ref{temp}c the normalized
superconducting peak ratio (SPR), which is defined after Ref.\ \cite{feng00} as
the ratio between the extracted peak intensity and the total spectral weight
(excluding the surface state peak) integrated over [0.3 eV, -0.1 eV]. Despite
the scatter in the data and the large error bars, the general trend suggests
that the appearance of this peak is related to the superconducting transition.

Furthermore, the remarkable resemblance of the spectra in Fig.\ \ref{scp}b
implies that this `peak-dip-hump' structure at X is a CuO$_2$ plane derived
feature since the CuO chain is absent in Bi2212. This argument is further
supported by data taken on Zn-doped ($\sim$2\%) Y123 with the same oxygen
content under identical experimental conditions (Fig.\ \ref{scp}c). Compared
with the data from Zn-free Y123 (Fig.\ \ref{scp}a), the major change is that
the superconducting peak shifts to lower BE, {\it i.e.}, the gap size decreases
upon Zn-doping. Since Zn atoms only occupy the Cu sites in the CuO$_2$ plane
\cite{xiao88}, this change in the superconducting peak suggests that it is
directly related to the CuO$_2$ plane. Due to the weak dispersion of this peak
around ($\pi,0$), we can approximately quantify the superconducting gap using
the peak position, giving a gap size of 25 and 29 meV at X for Zn-doped ($T_c$
= 79 K) and Zn-free ($T_c$ = 89 K) Y123, respectively. This indicates that, as
expected, the gap magnitude scales with $T_c$.

At this point, one may speculate as to why the `peak-dip-hump' feature is
absent at Y in Fig.\ \ref{edc}b. This is partly due to the presence of the
intense surface state peak, which masks the underlying bulk electronic states
close to $E_F$. In fact, the data taken at Y with the polarization
perpendicular to the chain direction (Fig.\ \ref{scp}d), where the surface
state peak is suppressed, clearly show a second peak at higher BE. In analogy
with the assignment of the superconducting peak at X we may reasonably
interpret this additional feature as evidence for a superconducting peak at Y.
The hump associated with this peak is almost indiscernible in Fig.\ \ref{scp}d,
but becomes more evident moving away from Y to S as it disperses towards $E_F$
(Fig.\ \ref{disp}f). The temperature dependent measurements on this feature are
very difficult due to its weak intensity and the presence of the strong surface
state peak. Hence, we deliberately aged the sample by warming it to the point
where the surface state peak completely vanishes (Fig.\ \ref{temp}d). A small
peak on top of a relatively large background exhibits a reproducible
temperature dependence expected for a superconducting peak. As a consequence of
aging, this peak shifts to lower BE ($\sim$33 meV), as compared with Fig.\
\ref{scp}d. Since the Y-point is outside the 1D chain FS, the peak and the hump
detected at Y should be exclusively related to the CuO$_2$ plane.

Having identified all of the features in the low energy excitation spectra, our
data provide the first observation of a `peak-dip-hump' structure near
($\pi,0$) in a system other than Bi2212. This finding partially resolves the
longstanding issue regarding the universality of the superconducting peak,
which is crucial for a complete understanding of HTSC as it appears to contain
information on both pairing and phase coherence \cite{feng00}. The failure to
detect this peak in previous photoemission studies on Y123 was mainly due to
the presence of the intense surface state peak and insufficient energy and
momentum resolution. However, it is not clear at present whether the absence of
this peak in other superconducting cuprates, such as
Bi$_2$Sr$_2$CuO$_{6+\delta}$ and La$_{2-x}$Sr$_x$CuO$_4$, is simply a
consequence of lower $T_c$ \cite{note1}, {\it i.e.}, low superfluid density, or
is also related to the lack of the CuO$_2$ bilayer that exists in both Bi2212
and Y123. Further experiments on other cuprate superconductors, particularly
the single-plane Tl- or Hg-based compounds with higher $T_c$, are necessary to
clarify this issue.

Now let us turn our attention to the strong $a$-$b$ anisotropy of the
electronic structure as revealed by a direct comparison between Fig.\
\ref{edc}a and \ref{edc}b. For the chain state, the $a$-$b$ asymmetry is in
accord with the 1D character of the CuO chain. More interesting is the $a$-$b$
anisotropy associated with the peak and the hump as they are related to the
superconductivity in the CuO$_2$ plane. The major asymmetry lies in the energy
positions of the peak and the hump: 29 ($\Delta_x$) and 120 meV ($\omega_x$)
for X, and 44 ($\Delta_y$) and 180 meV ($\omega_y$) for Y. This yields a
significant ($\sim$50\%) difference in the gap magnitude between X and Y, a
substantial deviation from the ideal $d$-wave pairing state. This $a$-$b$
asymmetry of the superconducting gap in Y123 has also been reported by Limonov
and co-workers in their Raman scattering study \cite{limonov00}, from which a
value of 25 and 30 meV can be derived for $\Delta_x$ and $\Delta_y$,
respectively. It is also remarkable to see that the gap magnitude scales with
the hump positions: $\Delta_x / \omega_x \approx \Delta_y / \omega_y \approx
0.24$. The scaling of these two energy scales has also been observed by ARPES
in Bi2212 as a function of doping \cite{white96,campuzano99}. The observation
of this scaling behavior in the same material along two different directions
imposes strong constraints on theory.

In addition to the asymmetry of the gap magnitude, we also notice that there is
a clear difference in the intensities of the peak and the hump between X and Y.
According to Ref.\ \cite{feng00}, this could possibly be a signature of the
in-plane anisotropy of the superfluid density. However, the polarization
effects due to the presence of the CuO chains on the surface layer may
complicate the situation, making a quantitative comparison difficult. We note
that strong $a$-$b$ anisotropy of the penetration depth ({\it i.e.}, the
superfluid density) was also found in microwave and far infrared spectroscopic
measurements \cite{zhang94,basov95}, which was interpreted as evidence for the
presence of superconductivity in the chain layer. While it is still a
controversial issue whether the ground state of the CuO chain is
superconducting or a charge density wave \cite{grevin00}, our data offer an
alternative interpretation: this anisotropy may in fact reside in the CuO$_2$
plane itself.

Such a strong $a$-$b$ anisotropy is crucial for a comprehensive understanding
of the superconducting properties of Y123 and the interpretation of related
experiments. It is therefore important to understand the origin of this
in-plane anisotropy. One possible explanation could be based on the
orthorhombicity of the CuO$_2$ plane in Y123. According to the LDA band
calculations, the ratio between the hopping integrals along the $a$ and $b$
directions scales roughly as $(b/a)^4$, which, in turn, yields considerable
differences in the calculated electronic structures between $\Gamma$-X-S and
$\Gamma$-Y-S \cite{anderson95} that qualitatively agree with our data. However,
no direct conclusion can be drawn for the superconducting state properties.
Alternatively, a strong chain-plane coupling may be present. In particular, as
recently suggested by Atkinson \cite{atkinson99}, a nontrivial $d$-wave gap
structure, which seems to be consistent with our data, exists on the
anti-bonding band related FS due to the hybridization with the chain band.

We thank for J. C. Davis, A. L. de Lozanne, H. Eisaki, T. Mizokawa and S.
Schuppler for fruitful discussions, T. Tohyama, and S. Maekawa for unpublished
results from their calculations. The data presented here were obtained at SSRL,
which is operated by the Department of Energy, Office of Basic Energy Sciences.
The Stanford work was also supported by ONR Grant N00014-98-1-0195 and NSF
grant DMR-9705210.


\begin{references}
\bibitem{note1} Even in the Bi2212 system, no superconducting peak has been resolved for samples with $T_c < 40$ K.
\end{references}
\end{document}